\documentclass{ws-p10x7}

\begin{document}

\title{$\varepsilon'/\varepsilon$ in the standard model with hadronic 
       matrix elements from the chiral quark model}

\author{M.\ Fabbrichesi}

\address{INFN and SISSA, via Beirut 4, I-34014 Trieste, 
         Italy\\E-mail: marco@he.sissa.it}

\twocolumn[\maketitle\abstract{I discuss the estimate
   of the CP-violating parameter $\varepsilon'/\varepsilon$ based on
   hadronic matrix elements computed in the chiral quark
   model. This estimate suggested, before
   the current experimental results, that the favored value of
   $\varepsilon'/\varepsilon$ in the standard model is of the order of
   $10^{-3}$. I briefly review the physical effects on which this
   result is based and summarize current estimates.}]

If we imagine to be back in 1997---looking at the experimental
results for the ratio  $\varepsilon'/\varepsilon$ and its
theoretical estimates---we will find ourselves in a rather confusing
situation in which the theoretical estimates favor
values of the order of $10^{-4}$ and the experiments disagree by more
than $3\sigma$ of their errors, and, moreover, do not rule out the
super-weak scenario in which  $\varepsilon'/\varepsilon$ vanishes (for
a review, see, \cite{review}). 

That was the situation when we decided to assess our theoretical
understanding and possibly provide a new estimate. 
The crucial point
was, and still is, that, if there is no sizable cancellation between
some of the relevant effective operators, the order of magnitude
of  $\varepsilon'/\varepsilon$ is bound to be of the order of
$10^{-3}$. A simple argument for this is presented
in~\cite{qcd99}. The problem is that any cancellation, or the lack thereof,
among the operators heavily 
depends on the size of the hadronic matrix elements and, in 1997,
there was no estimate of them that was free of hard-to-control
 assumptions.

Was it possible to improve on this situation?
We wanted to estimate the hadronic matrix elements in a systematic
manner without having first to solve QCD (not even by lattice
simulation).
To do this we needed a 
model that would be simple enough to understand its dynamics
and, 
at the same time, not too simple so as to
still include what we thought was the relevant physics. We chose the
chiral quark model~\cite{QM} in which all coefficients of the
relevant chiral lagrangian are parameterized in terms of just three 
parameters: the quark and gluon condensates, and the quark constituent
mass. The model makes possible a complete estimate of all matrix elements, it
includes non-factorizable effects, chiral corrections and final-state
interaction, all of which we thought to be relevant.
\begin{figure}[ht]
\epsfxsize180pt
\figurebox{180pt}{160pt}{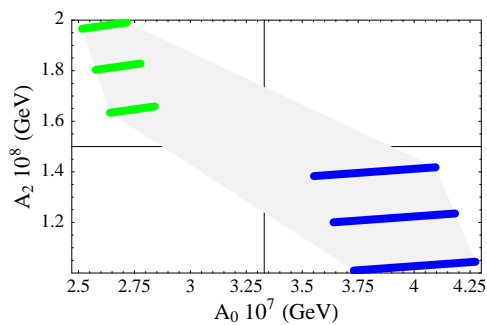}
\caption{Fitting the $\Delta I=1/2$ selection rule. The bars represent
values according to the given ranges of the model-dependent
parameters and other inputs. 
\label{fig:diHV}}
\end{figure}
In order to determine the three free parameters of the model, the
experimental CP-conserving, isospin $I=0$ and 2
components of the $K\to \pi \pi$ amplitudes, respectively 
$A_0$ and $A_2$, are fitted to obtain the values
reported in~\cite{trieste00} for the parameters. 
The  systematic uncertainty of this approach is
included  by varing the fit by 30\%
around the experimental values of the amplitudes.
Notice that the parameter values turn out to be rather close to those found by
independent estimates, even though {\it a priori} they could have been any
number. Moreover, the $\Delta I=1/2$ rule is reproduced in a 
natural manner (see~\cite{1/2} for a discussion). This rule is such a
fundamental feature of kaon physics that no estimate of
$\varepsilon'/\varepsilon$ can be said reliable unless it also
reproduces this selection rule.

These results are stable under changes of the renormalization scale
and $\gamma_5$-scheme (see \cite{trieste00} for details).

\begin{figure}[ht]
\epsfxsize180pt
\figurebox{180pt}{160pt}{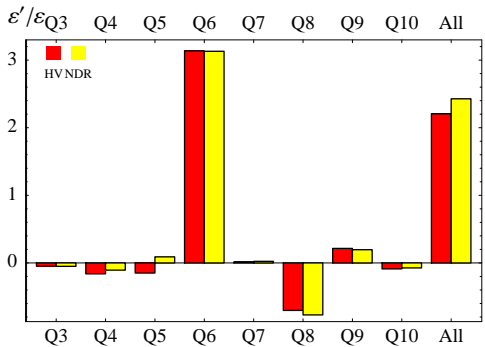}
\caption{Contribution of the various operators of the $\Delta S=1$
four-quark effective Hamiltonian to  $\varepsilon'/\varepsilon$.
\label{fig:histo}}
\end{figure}

\begin{figure}[ht]
\epsfxsize180pt
\figurebox{180pt}{160pt}{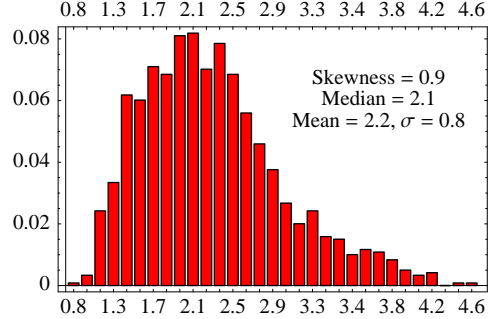}
\caption{Distribution of values of  $\varepsilon'/\varepsilon$ at the
varying of the input parameters. 
\label{fig:dist}}
\end{figure}

Having fixed the model-dependent parameters, we can proceed
and compute the ratio  $\varepsilon'/\varepsilon$. As it can been seen
from fig.~\ref{fig:histo}, the gluon penguin operator $Q_6$ dominates
all other operators so that the final value of
CP-violating ratio turns out to be of the expected order of $10^{-3}$,
and the standard model does not mimic the super-weak scenario. This is
the main result of our analysis; its publication in
1997~\cite{trieste97}
correctly predicted the current experimental results. The present
estimate is an update of the short-distance inputs which also contains an
improved treatment of the uncertainties. 
To estimate the uncertainty of our result
we can vary, according to a Gaussian
distribution, all the short-distance inputs and by a flat distribution the
model-dependent parameters to obtain the distribution of
values shown in fig.~\ref{fig:dist}.
Such a distribution gives the value
\begin{equation}
\varepsilon '/\varepsilon = (2.2 \pm 0.8) \times 10^{-3} \, ,
\label{eq:Gauss}
\end{equation}
in good agreement with the current experimental average~\cite{exps}
\begin{equation}
\varepsilon/\varepsilon' = (1.9 \pm 0.46) \times 10^{-3} \, ,
\label{eq:exps}
\end{equation}
where the error has been inflated according to the Particle Data Group
procedure to be used when 
averaging over experimental data with 
substantially different central values.
In a more conservative approach all inputs are varied with uniform
probability over their whole ranges to obtain
\begin{equation}
0.9 \times 10^{-3} < \varepsilon '/\varepsilon < 4.8 \times 10^{-3} \, .
\label{eq:flat}
\end{equation} 
Given the intrinsec difficulty of the computation, I do not 
expect in the near future smaller uncertainties.

It is easy to go back into the computation and understand the final
result. Chiral loops and final-state interactions both tends to
enhance the $A_0$ amplitudes by making the gluon penguin contribution
larger. Larger gluon penguins dominate the contribution of the
electro-weak sector in  $\varepsilon '/\varepsilon$ and no effective
cancellation between the two occurs. 
Non-factorizable (soft) gluon corrections make $A_2$
smaller. They play an important role in the $\Delta I=1/2$ rule and in
the determination of the model-dependent parameters although
not directly in $\varepsilon '/\varepsilon$ where only penguin operators
enter. Most of these effects can be summarized by saying that the bag
factor $B_6$ of the the gluon operator $Q_6$ is much larger  
(at a given scale) than its vacuum-saturation value of 1.

Many of the points suggested by the chiral quark model
analysis have been taken up by other groups after the current
experiments favored a value of  $\varepsilon '/\varepsilon$ of the
order of $10^{-3}$. In particular, chiral
corrections~\cite{1/N,dubna},
non-factorizable effects~\cite{cheng},
final-state interactions~\cite{Pich} and effective-model estimates~\cite{Hans}
have been discussed  recently. 

\begin{figure}[ht]
\epsfxsize180pt
\figurebox{180pt}{160pt}{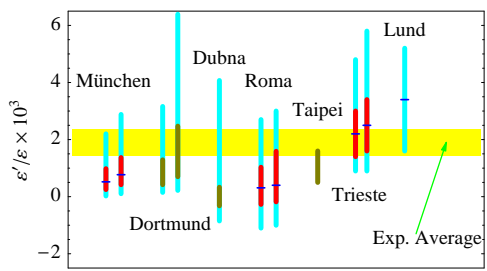}
\caption{Experiments vs.\ theoretical prediction in the year
2000. References are quoted in the text.
\label{fig:expvsth}}
\end{figure}

In Fig.~\ref{fig:expvsth} current 
estimates~\cite{trieste00,1/N,dubna,others,Hans} are summarized; the
same figure
shows that, nowadays, contrarily to
what is still too often repeated in papers and seminars,
most standard model estimates agree with the experiments and with the
prediction of the chiral quark model.

Because of its simplicity, the chiral quark
model is clearly not the final word
and it can now been abandoned---as a ladder used to climb a wall after we
are on the other side---as we work for better estimates, in
particular, those from the lattice simulations.

\section*{Acknowledgments}

It is a pleasure to thank my collaborators S.~Bertolini and J.~O.~Eeg
and my former students V.~Antonelli and E.~I.~Lashin 
for the work done together.

\end{document}